\documentclass[twocolumn]{aastex61}
\usepackage{amssymb}
\usepackage{multirow}
\usepackage{morefloats}
\usepackage{amsmath}
\usepackage{bigstrut}
\usepackage{booktabs}
\usepackage{natbib}
\usepackage{float}
\usepackage{graphicx,epsfig,fancyhdr,epsf,txfonts,epstopdf}
\usepackage{latexsym,bbm}
\usepackage{lineno}
\usepackage{color}
\usepackage{ulem}
\usepackage{stfloats}

\received{XXX}
\revised{YYY}
\accepted{ZZZ}

\shortauthors{Zhang et al.}

\begin{document}

\shorttitle{X-ray Variability of Mrk 421}

\title{X-ray Intra-day Variability of the TeV Blazar Mrk 421 with Suzaku}

\author{Zhongli Zhang}\thanks{Email: zzl@shao.ac.cn}
\affil{Shanghai Astronomical Observatory, Key Laboratory of Radio Astronomy, Chinese Academy of Sciences, 80 Nandan Road, Shanghai 200030, China}

\author{Alok C.\ Gupta}\thanks{Email: acgupta30@gmail.com}
\affiliation{Aryabhatta Research Institute of Observational Sciences (ARIES), Manora Peak, Nainital -- 263002, India}

\author{Haritma Gaur}\thanks{Email: harry.gaur31@gmail.com}
\affiliation{Aryabhatta Research Institute of Observational Sciences (ARIES), Manora Peak, Nainital -- 263002, India}

\author{Paul J. Wiita}
\affiliation{Department of Physics, The College of New Jersey, 2000 Pennington Rd., Ewing, NJ 08628-0718, USA}

\author{Tao An}
\affiliation{Shanghai Astronomical Observatory, Key Laboratory of Radio Astronomy, Chinese Academy of Sciences, 80 Nandan Road, Shanghai 200030, China}

\author{Minfeng Gu}
\affiliation{Key Laboratory for Research in Galaxies and Cosmology, Shanghai Astronomical Observatory, Chinese Academy of Sciences,
80 Nandan Road, Shanghai 200030, China}

\author{Dan Hu}
\affiliation{Department of Astronomy, Shanghai Jiao Tong University, 800 Dongchuan Road, Minhang, Shanghai 200240, China}

\author{Haiguang Xu}
\affiliation{School of Physics and Astronomy, Shanghai Jiao Tong University, 800 Dongchuan Road, Minhang, Shanghai 200240, China}
\affiliation{Tsung-Dao Lee Institute, Shanghai Jiao Tong University, 800 Dongchuan Road, Minhang, Shanghai 200240, China}
\affiliation{IFSA Collaborative Innovation Center, Shanghai Jiao Tong University, 800 Dongchuan Road, Minhang, Shanghai 200240, China}

\begin{abstract}
\noindent
We present X-ray flux and spectral analyses of the three pointed {\it Suzaku} observations of the TeV high 
synchrotron peak blazar Mrk 421 taken throughout its complete operational duration. The observation 
taken on 5 May 2008 is, at  364.6 kiloseconds (i.e., 101.3 hours), the longest and most evenly sampled continuous 
observation of this source, or any blazar, in the X-ray energy 0.8 -- 60 keV
until now. We found large amplitude 
intra-day variability in all soft and hard bands in all the light curves. The discrete correction function 
analysis of the light curves in soft and hard bands peaks on zero lag, showing that the emission in hard 
and soft bands are cospatial and emitted from the same population of leptons. The hardness ratio plots 
imply that the source is more variable in the harder bands compared 
to the softer bands. The source is
 harder-when-brighter, following the general behavior of high synchrotron peak blazars.
  Power spectral densities of all three light curves are red noise dominated, with a range of
power spectra slopes.  If one assumes that the emission originates very close to the
central super massive black hole, a crude  estimate for its mass, of $\sim 4 \times 10^{8}$ M$_{\odot}$, can be made;
 but if the variability is due to  perturbations arising there that are advected into the jet and are thus Doppler boosted, 
substantially higher masses are consistent with
the quickest seen variations.
We briefly discuss the possible physical mechanisms most likely responsible for the observed flux and spectral 
variability. 

\end{abstract}

\keywords{galaxies: active -- BL Lacertae objects: general -- quasars: individual -- BL Lacertae objects: 
individual: Mrk 421}

\section{Introduction}
\noindent
The blazar subclass of active galactic nuclei (AGNs) is usually taken to include BL Lacertae objects (BL Lacs) and
flat spectrum radio quasars (FSRQs). In their optical spectra, BL Lacs show either very weak (EW $< 5\AA$ ) or no emission 
lines \citep{Stocke1991,Marcha1996}, while FSRQs have strong emission lines \citep[e.g.,][]{Blandford1978,Ghisellini1997}. 
Blazars are primarily characterized by highly variable flux, high polarization in the radio to optical bands, 
core dominated radio structures, and emission being predominantly non-thermal across the entire electromagnetic (EM) 
spectrum. The emission is considered to mostly arise from the relativistic jet aligned at a small angle with observer's line of sight 
\citep[LOS; e.g.,][]{Urry1995}. \\ 
\\
The multi-wavelength emissions of blazers over the entire EM spectrum are characterized by 
broad double peaked structures in their spectral energy distributions (SEDs). The low energy peak of blazars 
SEDs lies between the infrared and X-ray bands and is a result of synchrotron emission from relativistic non-thermal 
electrons in the jet. The high energy component peaks in gamma-rays between GeV to TeV energies and probably originates 
from inverse Compton (IC) up-scattering of the synchrotron or external photons off the relativistic electrons in the jet
\citep[e.g,][]{Kirk1998,Gaur2010}. While these leptonic models usually seem to provide good fits to quasi-simultaneously measured
broad SEDs, hadronic models may be preferred in some cases \citep[e.g,][]{Diltz2015}. \\
\\
Blazars show flux variations on diverse timescales across the EM spectrum. The variability timescales range 
from a few minutes to years and even decades. Flux variations from minutes to less than a day are commonly known 
as intra-day variability (IDV) \citep{Wagner1995} or intra-night variability or micro-variability \citep{Goyal2012}. 
Changes occurring in intervals from  days to a few months are often called short term variability (STV) while flux changes over 
timescales of several months to years even decades are usually denoted as long term variability \citep[LTV;][]{Gupta2004}. 
In general, blazars' LTV and much of the STV across the entire EM spectrum can be well explained through the shock-in-jet 
model \citep[e.g.,][]{Marscher1985,Hughes1985}. \\
\\
Mrk 421 ($\alpha_{2000.0} =$ 11h 04m 27.2s, $\delta_{2000.0} =$+38$^{\circ}$12$^{'}$32$^{''}$) is a TeV blazar at 
redshift $z =$ 0.031. It was the first extragalactic object discovered at TeV energies with a detection at a significance 
of 6.3$\sigma$ made by the Whipple collaboration \citep{Punch1992}. Later it was confirmed by the {\it HEGRA} (high energy gamma 
ray astronomy) group \citep{Petry1996}. Since its discovery as TeV blazar, it has been extensively studied in X-ray 
and $\gamma-$ray energies \citep[e.g.,][and references therein]{Kerrick1995,Gaidos1996,Maraschi1999,Brinkmann2001,Aharonian2003,Massaro2004,Nicastro2005,Tramacere2009,Abdo2011,Pian2014,Isobe2015,Pandey2017,Aggrawal2018}. 
It is one of the most frequently studied blazars thanks  to its strong, rapid and peculiar 
flux variations throughout the EM spectrum and it has been the subject of several extended duration multi-wavelength observing 
campaigns \citep[e.g.,][and references therein]{Macomb1995,Tosti1998,Takahashi2000,Blazejowski2005, Rebillot2006,Fossati2008,Acciari2011,Aleksic2015,Bartoli2016,Ahnen2016}.\\ 

\begin{table*}
\centering
Table 1. The {\it Suzaku} observations of Mrk 421\\
\label{tab:observation}
\begin{tabular}{ccccccccccc}\hline\hline
 ObsID     & Date       & MJD &Elapse\tablenotemark{a} & Exp.\tablenotemark{b} & GTI\tablenotemark{c}  & Win.\tablenotemark{d}   & Snap\tablenotemark{e} & Rate 1\tablenotemark{f} &  Rate 2\tablenotemark{g}   &  Rate 3\tablenotemark{h}  \\ 
           &            &     & (ks) & (ks) & (ks)   &     & (s)     & (count s$^{-1}$) &  (count s$^{-1}$) &  (count s$^{-1}$) \\\hline
 701024010 & 2006-04-28 & 53853 &  82.0  &  41.5 &  31.9   & 1/4 & 2.0  & 41.2 &  42.2  & 1.59  \\
 703020010 & 2008-12-03 & 54591 & 190.0 &  101.3 & -- & 1/4 & 2.0  & 28.5 & -- &  0.77  \\
 703043010 & 2008-05-05 & 54803 & 364.6 &  180.8 & 146.5  & 1/8 & 1.0  & 37.0 &  37.5 &  0.91 \\\hline
\end{tabular}\\
\noindent
\tablenotemark{a}{Total elapsed time of the observation.}\\
\tablenotemark{b}{Total exposure time of the observation.}\\
\tablenotemark{c}{Common good time interval (GTI) of XIS and HXD/PIN applied for Obs. 701024010 and 703043010.}\\
\tablenotemark{d}{XIS window mode (1/4 or 1/8).}\\
\tablenotemark{e}{XIS snap time (time resolution).}\\
\tablenotemark{f}{Mean XIS 0 CCD count rate across the whole energy band.}\\
\tablenotemark{g}{Mean XIS 0 CCD count rate across the whole energy band after filtering common GTI of XIS and PIN.}\\
\tablenotemark{h}{The cleaned HXD/PIN count rate across the whole energy band after filtering common GTI of PIN and NXB.}\\
\end{table*}

\noindent
From Mrk 421, gamma-ray radiation has been observed in the energy range 50 MeV to 1 GeV by {\it EGRET} 
(Energetic Gamma Ray Experiment Telescope) onboard the Compton Gamma-ray Observatory; this 
was the first detection of gamma-ray emission from a BL Lac made by {\it EGRET} \citep{Lin1992}. 
\citet{Kerrick1995} reported a gamma-ray flare in the blazar on 1994 May 14 and 15 which showed an 
increase in flux by a factor of $\sim$10 compared to the quiescent level. After one day of this gamma-ray 
flare, a continuous 24 hours observations by {\it ASCA} showed the X-ray flux in a very high state. The 
2--10 keV flux peaked at 3.7 $\times$ 10$^{-10}$ ergs cm$^{-2}$ s$^{-1}$ and then decreased to 
1.8 $\times$ 10$^{-10}$ ergs cm$^{-2}$ s$^{-1}$ demonstrating large X-ray IDV \citep{Takahashi1996}. 
In May 1996, the Whipple telescope recorded two dramatic TeV outbursts from Mrk 421. The first outburst, 
with a doubling time of around one hour, showed the flux increased by a factor of $\sim$50 relative to the quiescent 
value, while in the second outburst, which lasted about 30 minutes,  the flux increased by a 
factor of 20--25 \citep{Gaidos1996}. Both outbursts showed strong IDV. A coordinated observation made in X-rays by
  {\it BeppoSAX} and $\gamma$-rays by  the Whipple telescope, in 1998 April gave the first 
evidence that the X-ray and TeV intensities are well correlated on timescales of hours \citep{Maraschi1999}.\\
\\
Mrk 421 was observed on several occasions during 2000 -- 2004 with {\it XMM-Newton} and was found to be in 
different flux states, i.e., stable, declining, and rising, and it often showed large amplitude 
IDV \citep{Brinkmann2001,Brinkmann2003,Brinkmann2005,Ravasio2004}. \citet{Cui2004} reported Mrk 421 
observations from {\it RXTE} and detected large amplitude IDV on several occasions. In 2000 February and May, 
X-ray and gamma-ray coordinated observations by {\it RXTE} and {\it HEGRA}, respectively, were done for Mrk 421. 
In both the energies rapid flux variabilities with different variability timescales were 
seen \citep{Krawczynski2001}. In 2013 April observations of Mrk 421 with {\it NuStar} in harder X-rays 
(3--79 keV) showed  large amplitude IDV in the blazar \citep{Paliya2015}. Recently, using
all the public archive data of {\it NuStar} and {\it Chandra}.  detailed IDV studies of Mrk 421
were carried out which show large amplitude IDV detections on several occasions, during which the soft and hard X-ray 
bands were well correlated \citep{Pandey2017,Aggrawal2018}. \\ 
\\
Brightness changes on the IDV timescales have been detected in large number of blazars in different 
EM bands \citep[e.g.,][and references therein]{Miller1989,Heidt1996,Sagar1999,Montagni2006,Aharonian2007,Gupta2008,Gaur2010,Gaur2012a,Kalita2015,Pandey2017,Paliya2017,Aggrawal2018}.
In general, blazar IDV observations carried out in different EM bands mentioned in the above 
papers last only for a few hours. In most of the cases the observations are not evenly sampled. 
But there are a few earlier observations in which IDV was examined over more extended periods of time in 
different EM bands \citep{Tanihata2001,Edelson2013}. \citet{Tanihata2001} used uninterrupted, long 
lasting ($\sim$ 7, 10, and 10 days, respectively) {\it ASCA} observations of three TeV blazars namely 
Mrk 421, Mrk 501, and PKS 2155$-$304 to study X-ray timing properties in the energy range 0.6 -- 2 keV 
and 2 -- 10 keV. Strong multiple flarings were detected in all the blazars during their observations
in both the  0.6 -- 2 keV and 2 -- 10 keV bands. The best cadence (30 minutes), nearly continuous and 
longest IDV observation of a blazar were done in the optical band by {\it Kepler} on W2R 1926$+$42 
\citep{Edelson2013}. Strong flux variation with multiple flares were seen and the flux distribution is highly 
skewed and non-Gaussian. {\it Kepler} data presented in \citet{Edelson2013} of the blazar W2R 1926$+$42 
were also used to study detailed variability and flare properties \citep{Bachev2015,Mohan2016,Sasada2017,Li2018}. 
These papers also used additional {\it Kepler} observations of that blazar. \\ 
\\    
This IDV is one of the most puzzling issues in blazar physics and may be related to  the  innermost 
region of activity close
to the central super massive black hole (SMBH).  IDV can certainly help constrain the 
 the size of emitting region and perhaps even the mass of the central SMBHs of blazars. 
However, performing such studies demands high cadence data for extended periods of time which is extremely 
difficult to obtain. \\
\\
In the present study, we are using all the public archived observations (i.e., three pointed observations) of Mrk 421 
which were observed by the {\it Suzaku} satellite during its period of operation.  
These three observations were carried out for 82.0, 190.0, and 364.6 ks on 2006 April 28, 2008 May 5, and
2008 December 3, respectively. These are quite long duration observations, and to the best 
of our knowledge, the 364.6 ks observation is the longest nominally continuous observation of any blazar 
in the broad X-ray energy band of 0.8 -- 60 keV with 
which one can study IDV with excellent data sampling. So, the data analyzed in this paper provides one of 
the best opportunities to understand the X-ray IDV behavior of one of the most interesting 
and peculiar blazars, Mrk 421. \\      
\\
The paper is structured as follows. In Section 2, we discuss the {\it Suzaku} public archival data of 
the blazar Mrk 421 used here, and its reduction. Section 3 gives information about the various analysis 
techniques used in the work. In Section 4 we present results and give a discussion of them in Section 5. 
Our conclusions are summarized in Section 6.           

\begin{figure*}
\begin{center}
\epsscale{1.0}
\includegraphics[width=8.9cm,angle=0]{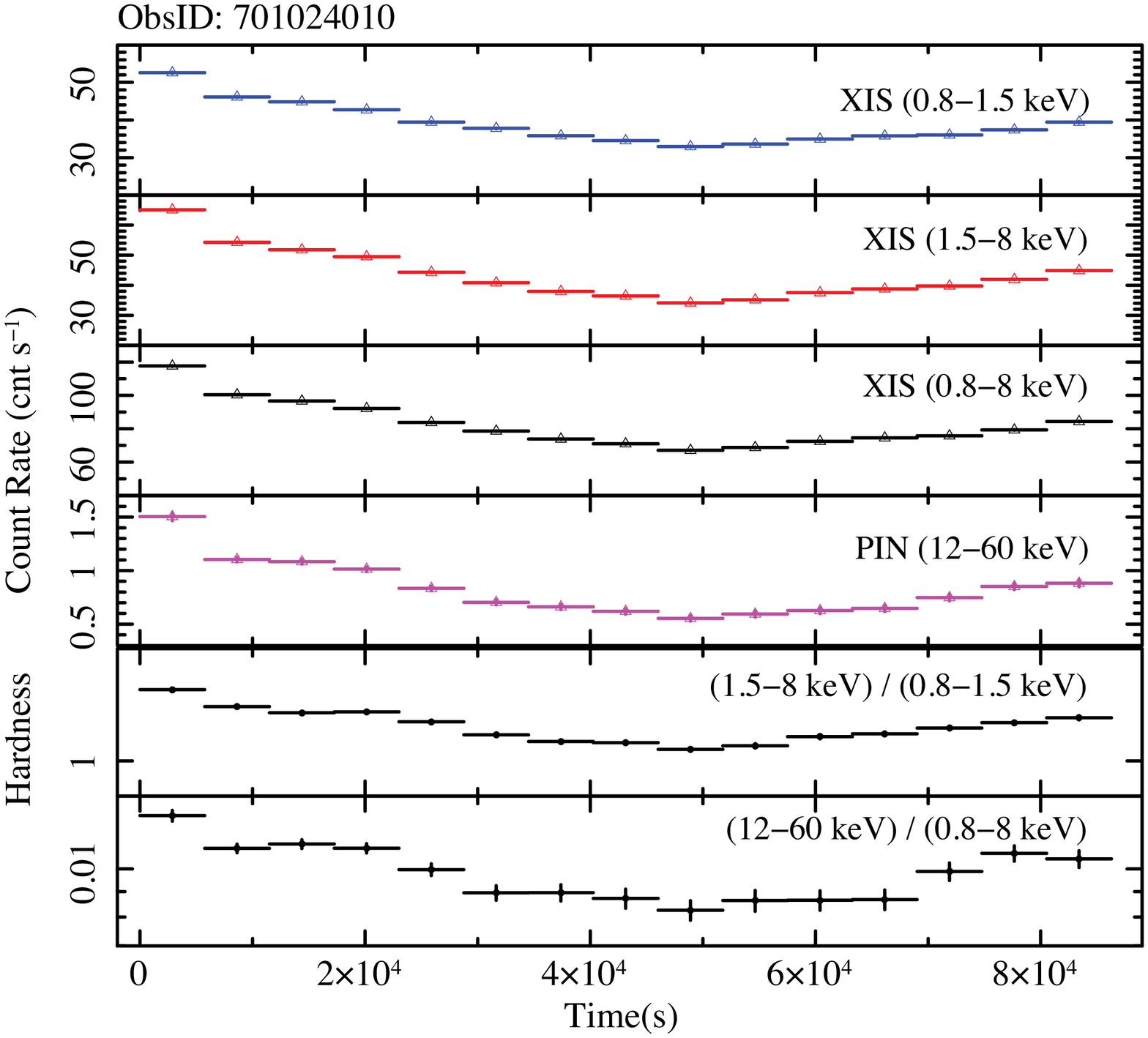} 
\includegraphics[width=8.9cm,angle=0]{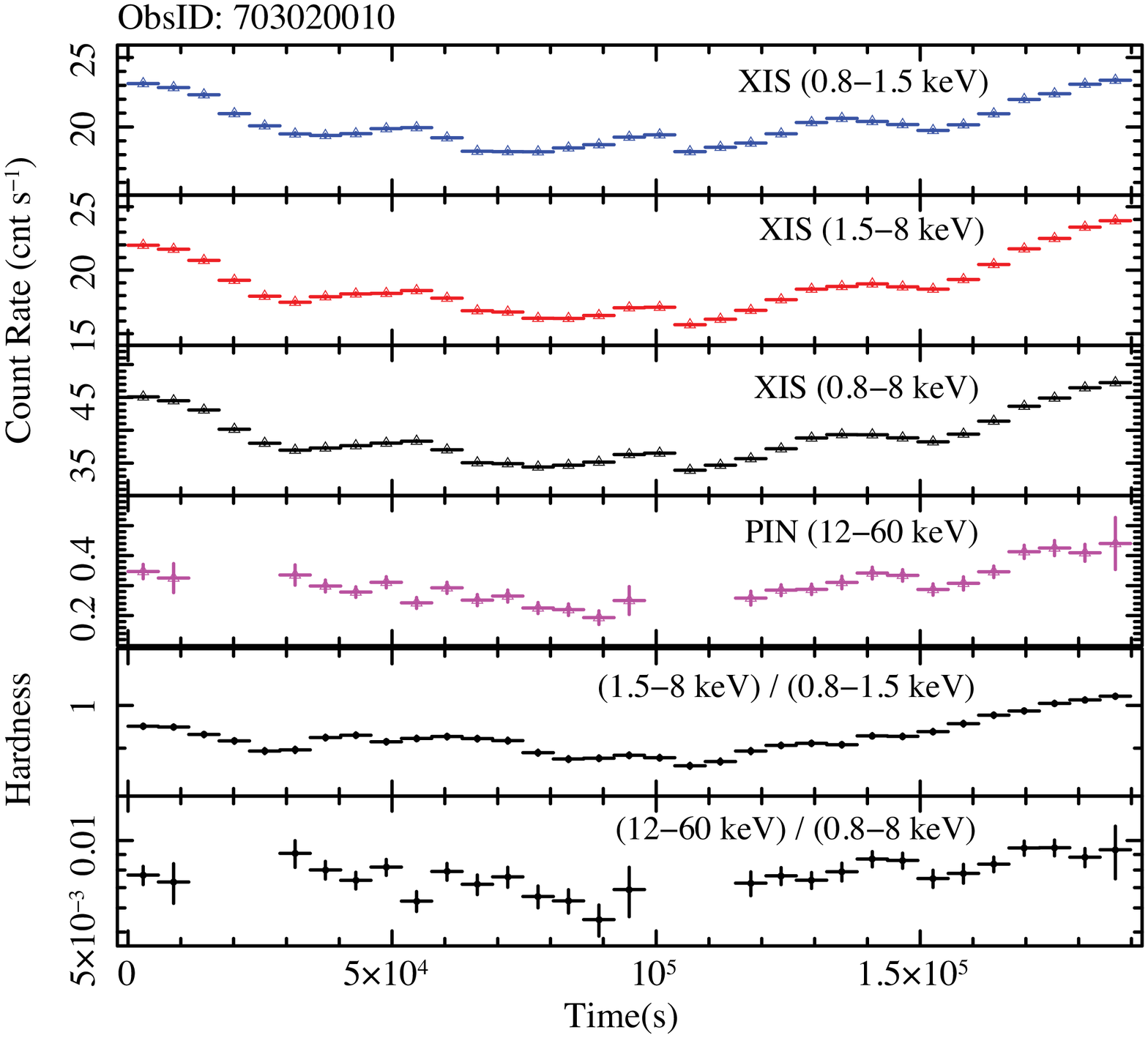}
\includegraphics[width=8.9cm,angle=0]{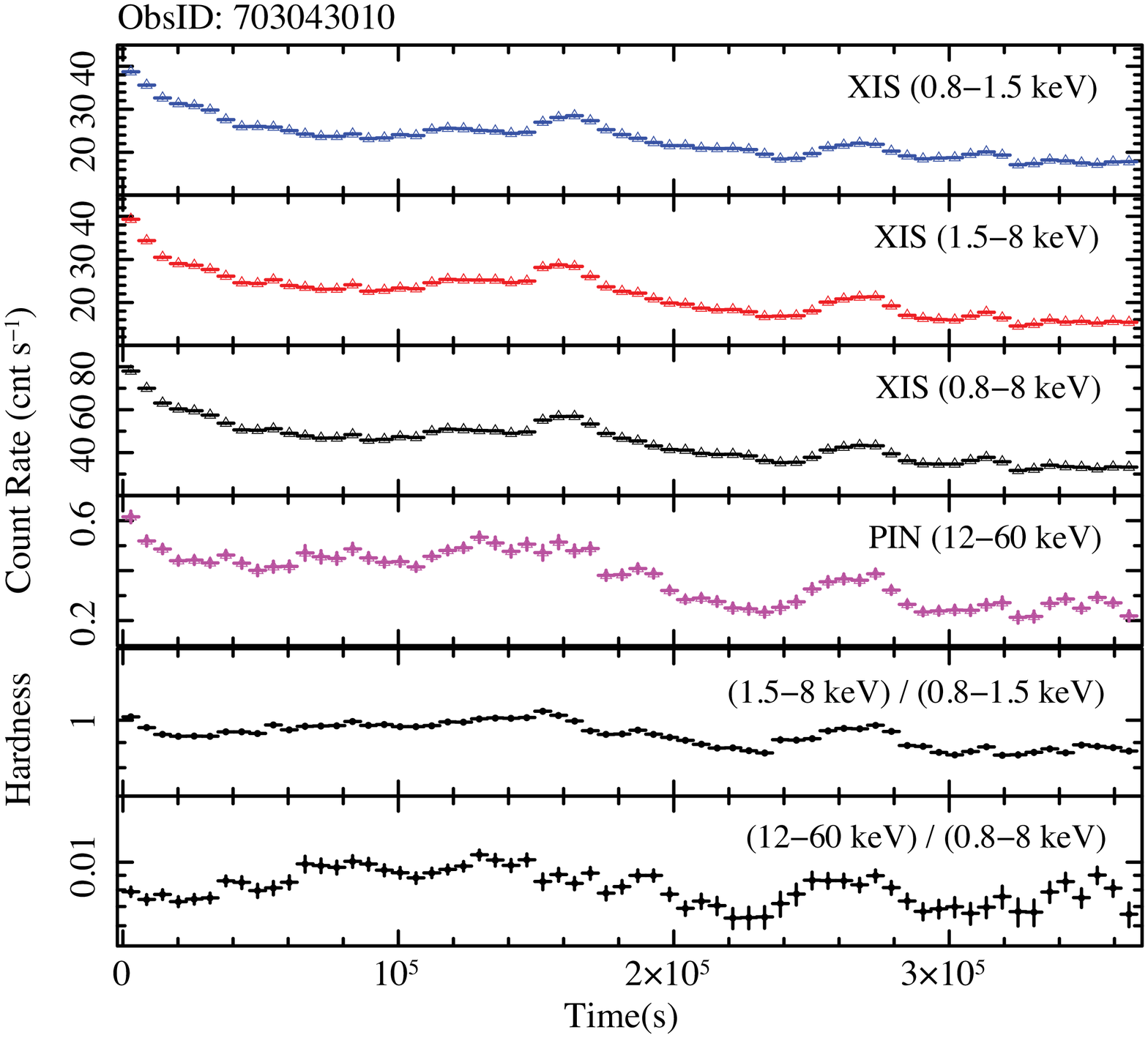} 
\figcaption{Light curves and Hardness Ratios of the three observations. The XIS soft (0.8$-$1.5 keV) and hard (1.5$-$8 keV) LCs are in blue and red, while the full XIS (0.8$-$8 keV) LC is in black.
The PIN (12$-$60 keV) LC is in magenta. For Obs 703020010, two big gaps are present around 20 and 110 ksec. The hardness ratios roughly follow the fluctuations of the LCs. }
\label{fig:count}
\end{center}
\end{figure*}

\section{{\it Suzaku} Archival Data Reduction}
\label{sec:observation}
\noindent
The Japanese X-ray observatory {\it Suzaku} is a near-earth satellite with an orbit apogee of 568 km and 
orbital period of 5752 s, which is shorter than the exposure request of many X-ray observations, including those
discussed here. As a result, 
most targets will be occulted by the Earth for $\sim1/3$ of each orbit. Additionally, the interruption of data acquisition by passages through the South 
Atlantic Anomaly (SAA) means that observing efficiency of the satellite is normally $< 50\%$. However, {\it Suzaku} retains excellent X-ray 
sensitivity with a broad-band energy range of 0.2 -- 60.0 keV, which makes it ideal to study objects with  high 
energy emission \citep{Mitsuda2007}. \\
\\
{\it Suzaku} consists of: the X-ray Imaging Spectrometer \citep[XIS,][]{Koyama2007} for the 0.2$-$10 keV low 
energy band;  the Hard X-ray Detector \citep[HXD,][]{Takahashi2007,Kokubun2007}, which utilizes PIN silicon 
diodes and is sensitive over 12$-$60 keV; and the GSO scintillator, which extends the detection ability to 
hundreds of keV. Mrk 421 was observed three times by {\it Suzaku} on 2006 April 28 (ID 701024010), 2008 May 5 
(ID 703043010) and 2008 December 3 (ID 703020010) \citep{Ushio2009,Ushio2010,Garson2010}. The 
observations are listed in Table \ref{tab:observation} in order of the ObsID and the total elapsed 
time. The longest observation (ID 703043010) lasted $\sim100$ hours, which is also the longest 
X-ray observation of this source by far. All observations were nominated on ``HXD" with ``normal" 
clock mode. The source was very bright during all three observations with XIS 0 CCD count rate of 
$\sim30-40$ count s$^{-1}$ (Table \ref{tab:observation}). In order to reduce pile-up effects, all 
XIS sensors were operated in 1/4 or 1/8 window modes, with short time resolutions of 2 or 1 seconds, 
respectively. Since the data were not significantly detected by HXD/GSO, we focused only on XIS 
and HXD/PIN data in this study. \\
\\
We processed the data with the  HEAsoft \citep[v6.17; HEASARC 2014;][]{Blackburn1995} software, with the calibration 
databases of version 20151005 for the XIS and version 20110913 for the HXD/PIN. We reprocessed the XIS data of 
all observations with the command {\it aepipeline}, while there is no necessity to have the PIN data reprocessed. ``Cleaned 
event files" were analyzed with screening of standard event selections. We utilized only front-illuminated 
(FI) XIS CCDs, which are XIS 0 and 3 for IDs 703020010 and 703043010, and XIS 0, 2 and 3 for ID 701024010, 
because they are more accurately calibrated than XIS 1 which uses a back-illuminated (BI) CCD chip. 
We excluded the central $0^{\prime}.5$ of all XIS CCDs to control the pile-up effect to under 3\% based 
on \citet{Yamada2012}. Hence the source regions were extracted from $0^{\prime}.5$ to $3^{\prime}$ which are 
consistent with \citet{Ushio2009,Garson2010} and \citet{Ushio2010}.  The source is so bright that the source signal
may contaminate the whole CCD. We extracted circular regions from the far edge of the XIS CCDs as the ``background", 
and found that the rates are $\sim$0.6\%, $\sim$0.8\% and $\sim$1.1\% of the source signals in the different observations; 
these are negligible, and thus were not subtracted in the further analysis.\\
\\
The cleaned HXD/PIN data were filtered mainly by conditions that the time intervals after exiting from the South 
Atlantic Anomaly should be longer than 500 sec, the elevation of the target above the Earth limb should be larger 
than 5$^{\circ}$, and the geomagnetic cutoff rigidity should be higher than 8 GV. We further screened the dead time of 
the cleaned PIN data with the command {\it hxddtcor}, and then applied common Good Time Intervals (GTIs) of PIN 
and its Non X-ray Background (NXB) using the command {\it mgtime}. The resulted PIN count rates of the three 
observations are listed in Table \ref{tab:observation}. We then subtracted the NXB according to the NXB model provided 
by the HXD team \citep{Fukazawa2009}, which take $\sim$35\%, $\sim$60\% and $\sim$ 61\% (with a systematic 
error $\lesssim$3\% according to \citet{Fukazawa2009}) of the cleaned PIN count rates of the three observations in 
Table \ref{tab:observation}. The contribution from the Cosmic X-ray Background (CXB) is constant in a certain energy 
band over time. Applying a model \citep{Boldt87},  
\begin{equation}
\textmd{CXB}(E) = 9.41\times 10^{-3}  \left(\frac{E}{1~\textmd{keV}}\right)^{\hspace{-0.3em}-1.29}\hspace{-1.4em}\exp\left(-\frac{E}{40~\textmd{keV}}\right)   
\label{eq:cxb}
\end{equation}
where the unit is photons cm$^{-2}$ s$^{-1}$ keV$^{-1}$ FOV$^{-1}$, the CXB was calculated to be 
$0.02$ cts s$^{-1}$, which is $<$ 3\% of the cleaned PIN count rates of the three observations. The model-dependent 
CXB level is negligible compared to the predominant NXB, and thus need not be subtracted from the light curves (LCs).\\
\\
As a standard procedure to compare observations of different energy bands, we extracted common Good Time 
Intervals (GTIs) of XIS and the cleaned HXD/PIN data using the command {\it mgtime}, resulting in the GTIs 
given in Table \ref{tab:observation}. For Obs. 703020010, good PIN data were not available in 2 out of 11 orbits of {\it Suzaku} 
at around $\sim20$ and $\sim110$ ksec, hence we could not apply a common GTI of XIS and PIN for this observation. 
The common GTIs contain 76.9\% and 81.0\% of the exposure times of Obs. 701024010 and 703043010, respectively. 
The amount of common GTI exposure times are mainly because of the way the HXD data were cleaned, as discussed in the above paragraph. 
The fluctuations of the XIS count rates of Obs. 701024010 and 703043010 after screening are less than 3\% as shown 
in Table \ref{tab:observation}, hence are negligible. We defined four sensitive energy bands as A (0.8$-$1.5 keV of XIS (soft)), 
B (1.5$-$8 keV of XIS (hard)), C (0.8$-$8.0 of XIS (total)) and D (12$-$60 keV of PIN (total)), and extracted corresponding LCs 
and hardness ratios as shown in Figure 1. The time binning was set to be exactly the orbital period (5752s) of {\it Suzaku} 
to most evenly sample the source in time; this makes for the most homogenous GTI fraction in each bin with statistically 
adequate counts. These span the temporal ranges of $\sim$30\% -- 60\% for the XIS data before applying common GTI with PIN, 
and become $\sim$20\% -- 60\% for the cleaned PIN data in each orbit of the three observations. The range of the GTI fraction per bin is 
mainly caused by the interruption of the SAA, which  obviously varies on timescales of an hour because of the earth's spin. However, 
we consider any discrepancy arising from this to be negligible because the source did not show large intrinsic variation within one orbit of Suzaku.
The light curves are essentially continuous, except for the PIN LC of Obs.\ 703020010, where the two gaps noted above 
are clearly seen. \\

\section{Analysis Techniques}
\label{sec:analysis}

\subsection{Excess variance}

\begin{table*}
Table 2. X-ray variability parameters
\centering
 \label{tab:var_par}
 \begin{tabular}{ccccccc} \hline \hline
          & \multicolumn{4}{c} { $F_{var}(percent)$}   &   &        \\\cline{2-5}
          & \multicolumn{3}{c} {XIS} & PIN             &  \multicolumn{2}{c} {$\tau_{var}$ (ks)}       \\\cline{2-4} \cline{6-7}
Observation  & Soft              & Hard            & Total           & Total           & XIS Total       & PIN Total       \\
ID           & (0.8 $-$ 1.5 keV) & (1.5 $-$ 8 keV) & (0.8 $-$ 8 keV) & (12 $-$ 60 keV) & (0.8 $-$ 8 keV) & (12 $-$ 60 keV)  \\\hline
701024010 &  10.81$\pm$0.10         & 14.90$\pm$0.09       &  12.95$\pm$0.07        & 30.99$\pm$0.86     & 36.37$\pm$0.60 & 18.58$\pm$1.92 \\
703020010 & ~~7.72$\pm$0.07         & 11.75$\pm$0.07       & ~~9.60$\pm$0.05        & 20.62$\pm$0.02     & 78.07$\pm$4.66 & 23.03$\pm$8.89 \\
703043010 &  19.53$\pm$0.06         & 23.43$\pm$0.06       &  21.34$\pm$0.04        & 27.02$\pm$0.70     & 47.16$\pm$2.49 & 23.16$\pm$6.68 \\\hline
\end{tabular}
\end{table*}

\noindent
Blazars show rapid and strong flux variations on diverse timescales across the EM spectrum. To quantify the 
strength of the variability, excess variance $\sigma_{XS}$, and fractional rms variability amplitude $F_{var}$ \citep[e.g.,][]{Edelson2002}, 
are often calculated. Excess variance is a measure of source's intrinsic variance determined by removing the variance 
arising from measurement errors from the total variance of the observed LC. If a LC consisting of $N$ measured flux values $x_i$, 
with corresponding finite uncertainties $\sigma_{err,i}$ arising from measurement errors, then the excess variance 
is calculated as
\begin{equation}
\sigma_{XS}^2 = S^2 - \bar\sigma_{err}^2
\end{equation}
where $\bar{\sigma_{err}^2}$ is the mean square error, defined as
\begin{equation}
\bar\sigma_{err}^2 =\frac{1}{N} \sum\limits_{i=1}^N \sigma^2_{err,i} 
\end{equation}
and $S^2$ is the sample variance of the LC, given by
\begin{equation}
S^2 = \frac{1}{N-1} \sum\limits_{i=1}^N (x_i - \bar{x})^2 
\end{equation}
where $\bar{x}$ is the arithmetic mean of $x_i$. \\

The normalized excess variance is $ \sigma^2_{NXS} = {\sigma^2_{XS}} / {\bar{x}^2}$
and the fractional rms variability amplitude, $F_{var}$, which is the square root of $\sigma^2_{NXS}$ is thus
\begin{equation}
F_{var} = \sqrt{\frac{S^2 - \bar{\sigma}^2_{err}}{{\bar{x}^2}}} 
\end{equation}
The uncertainty on $F_{var}$ is given by \citet{Vaughan2003}
\begin{equation}
err(F_{var}) =  \sqrt{\left( \sqrt{\frac{1}{2N}}\frac{\bar\sigma_{err}^2}{\bar{x}^2 F_{var}} \right)  ^ 2+ \left(  \sqrt{\frac{\bar\sigma^2_{err}}{N}} \frac{1}{\bar{x}}\right) ^2 }
\end{equation}

\subsection{Flux Variability Timescale}

\noindent
For variability timescale estimation, we followed the method described in \citet{Bhatta2018} which we also briefly 
describe here. 
According to \citet[][]{Burbidge1974}, a flux normalized, or weighted, variability timescale can be estimated by the following equation

\begin{equation}
   \tau_{var}=  \left | \frac{\Delta t}{ \Delta lnF} \right |
  \end{equation}

\noindent
where $\Delta t$ is the time interval between variable flux $F$ measurements  \citep[see also][]{Hagen-Thorn2008}. To compute the uncertainties in $\tau_{var}$, we used the standard error propagation method for a general function $y=f(x_{1}, x_{2}, . . x_{n})$ with the corresponding uncertainties $\Delta x_{1}, \Delta x_{2},  . . \Delta x_{n}$ in $x_{1}, x_{2}, . . x_{n}$, respectively. The uncertainties in $y$ can be expressed as \citep[similar to Equation 3.14 given in][]{Bevington2003}

\begin{equation}
\Delta y\simeq \sqrt{\left ( \frac{\partial y }{\partial x_{1}} \Delta x_{1} \right )^{2} +\left ( \frac{\partial y }{\partial x_{2}} \Delta x_{2} \right )^{2}+...+\left ( \frac{\partial y }{\partial x_{n}} \Delta x_{n} \right )^{2}
}
\label{error_prop}
\end{equation}

\noindent
Hence, by using Equation  \ref{error_prop}, uncertainties in $\tau_{var}$ are estimated as
\begin{equation}
\Delta \tau _{var}
\simeq \sqrt{\frac{F_{1}^{2} \Delta F_{2}^{2} +F_{2}^{2} \Delta F_{1}^{2}}{F_{1}^{2}F_{2}^{2}\left (  ln \left [ F_{1}/F_{2} \right ]\right )^{4}}}\ \Delta t
\end{equation}

\noindent 
Here $F_{1}$ and $F_{2}$ are the count rates (fluxes) used to estimate the shortest variability timescales, and $\Delta F_{1}$ and $ \Delta F_{2}$ are their corresponding uncertainties. \\
\\
These quantities characterizing flux variability in the blazar Mrk 421, i.e. fractional rms variability, and the weighted variability timescales are listed in columns 2 -- 7 of the Table 2, along with their errors. For estimating the weighted variability timescales, we used the LCs from 0.8 -- 8.0 keV (XIS total) and 12 -- 60 keV (PIN total). We note that the XIS $\tau_{var}$ values for the first and third observations are basically consistent ($\sim 40$ ks) but that for the middle one is roughly twice.  All three PIN  $\tau_{var}$ estimates for the higher energy variability are consistent, at around 20ks.  These faster weighted timescales can be understood in terms of the lower count rates at higher energies and changes in the hardness ratios described below, even though the overall XIS and PIN light curves are quite similar.

\subsection{Hardness Ratio}

\noindent
To characterize spectral variations of X-ray emission, the hardness ratio (HR) is an effective model independent tool. 
The hardness ratio is defined as 

\begin{equation}
HR = {\frac {H} {S}}
\end{equation}       

\noindent
where H and S are the net count rate in the hard, and soft energy bands, respectively. To study the spectral variability of Mrk 421 with {\it Suzaku}, we divided the XIS instrument energy into 0.8--1.5 keV (soft) and 1.5--8.0 keV (hard) 
bands.  We used the total energy of the XIS instrument 0.8--8.0 (soft) and PIN instrument total energy 12--60 keV as (hard) as our
other hardness ratio analysis. For both measurements of HRs we see that the variations are more pronounced at higher energies.

\subsection{Discrete Correlation Function}

\begin{figure*}
   \centering
\includegraphics[width=7cm , angle=0]{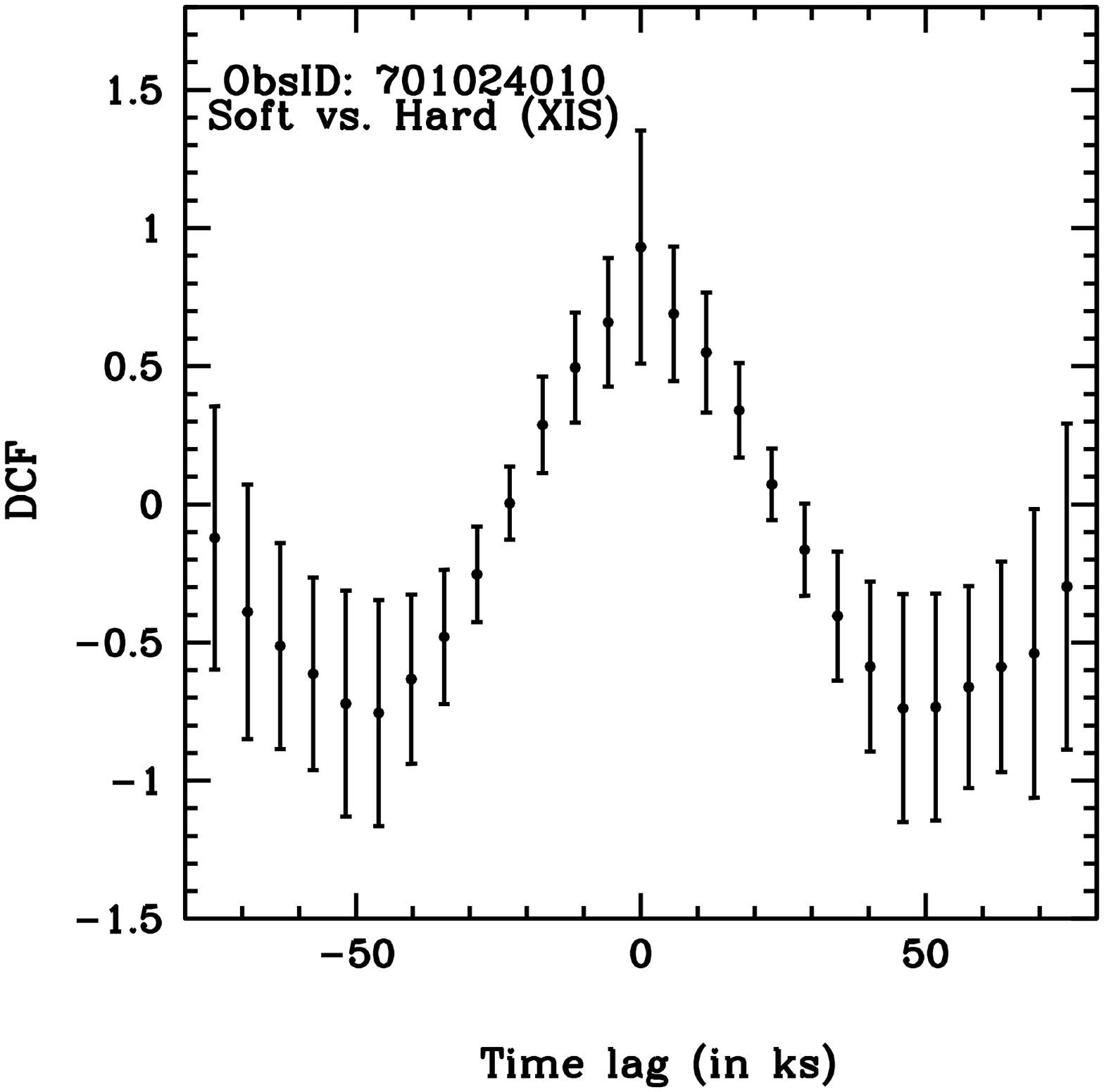}
\hspace*{0.7in} \includegraphics[width=7cm , angle=0]{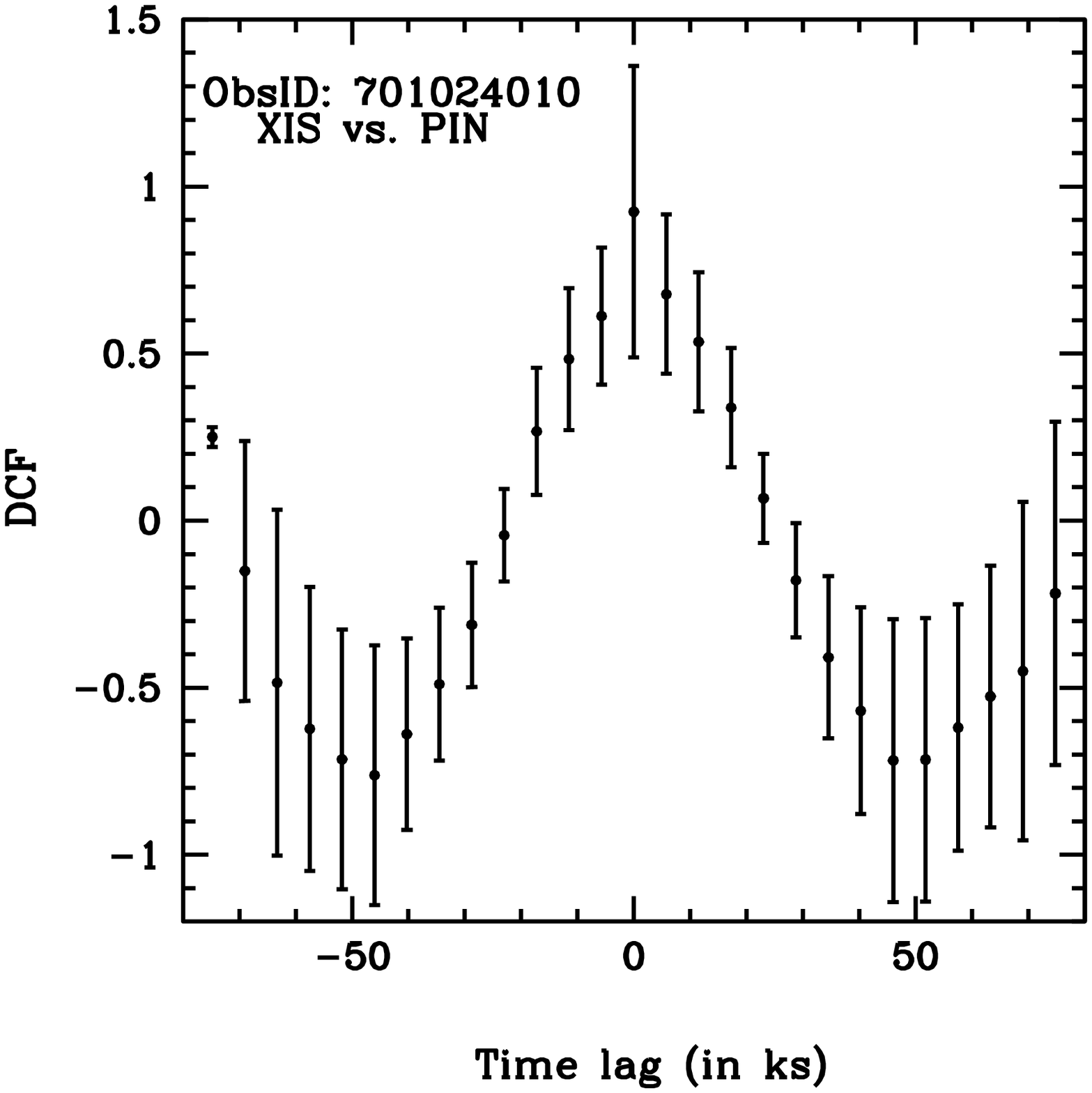}
\includegraphics[width=7cm , angle=0]{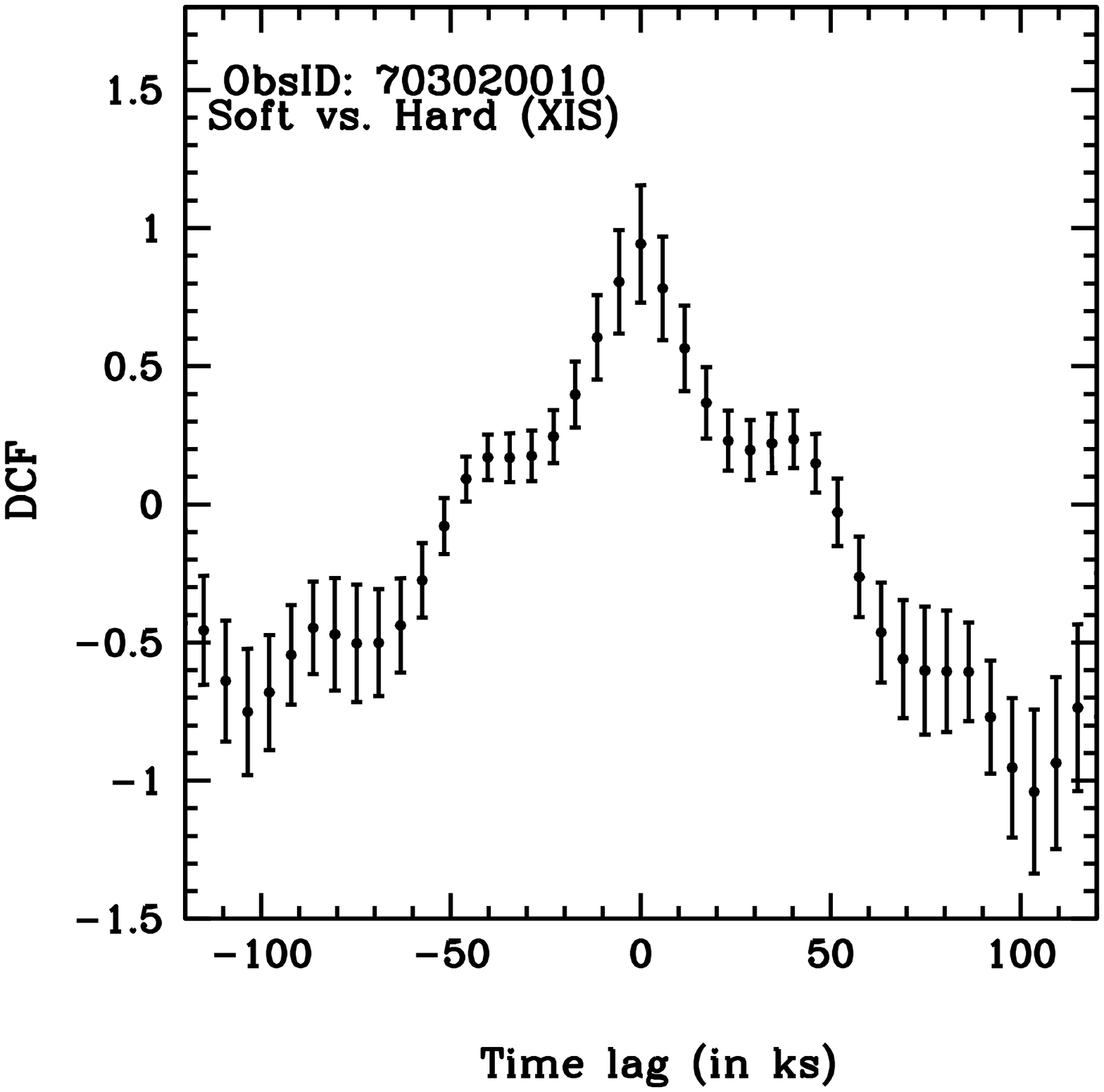}
\hspace*{0.7in} \includegraphics[width=7cm , angle=0]{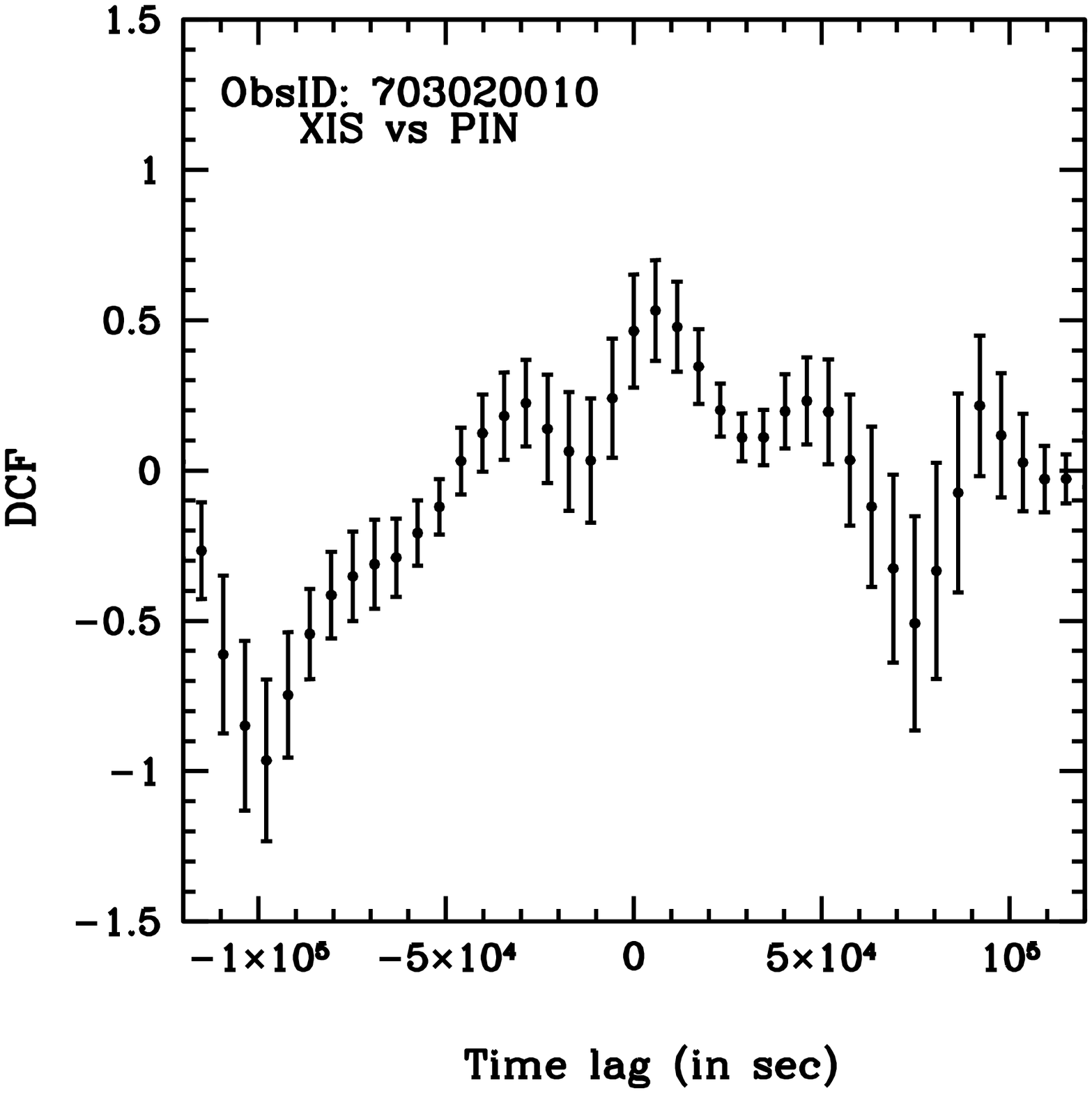}
\includegraphics[width=7cm , angle=0]{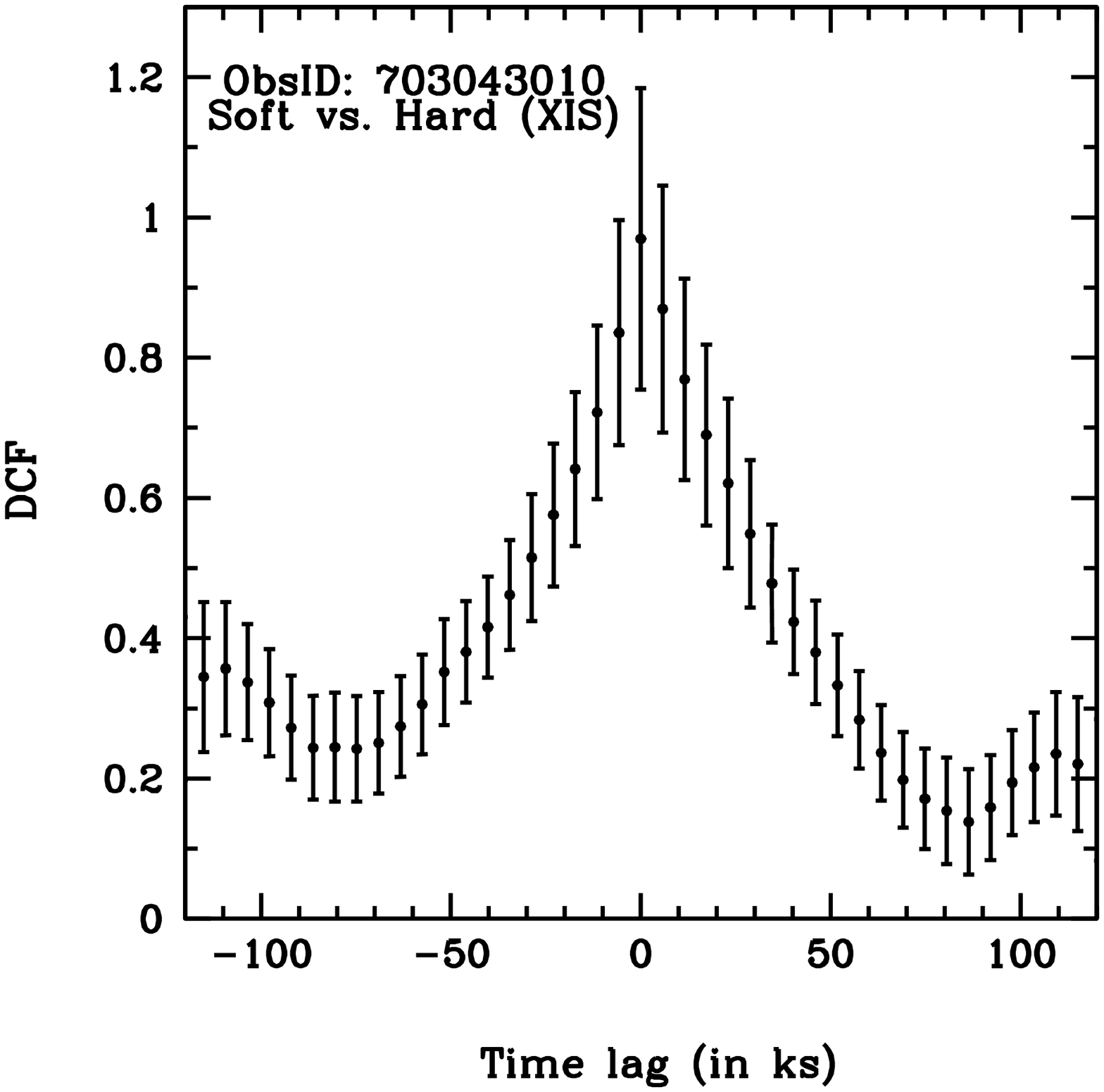}
\hspace*{0.7in} \includegraphics[width=7cm , angle=0]{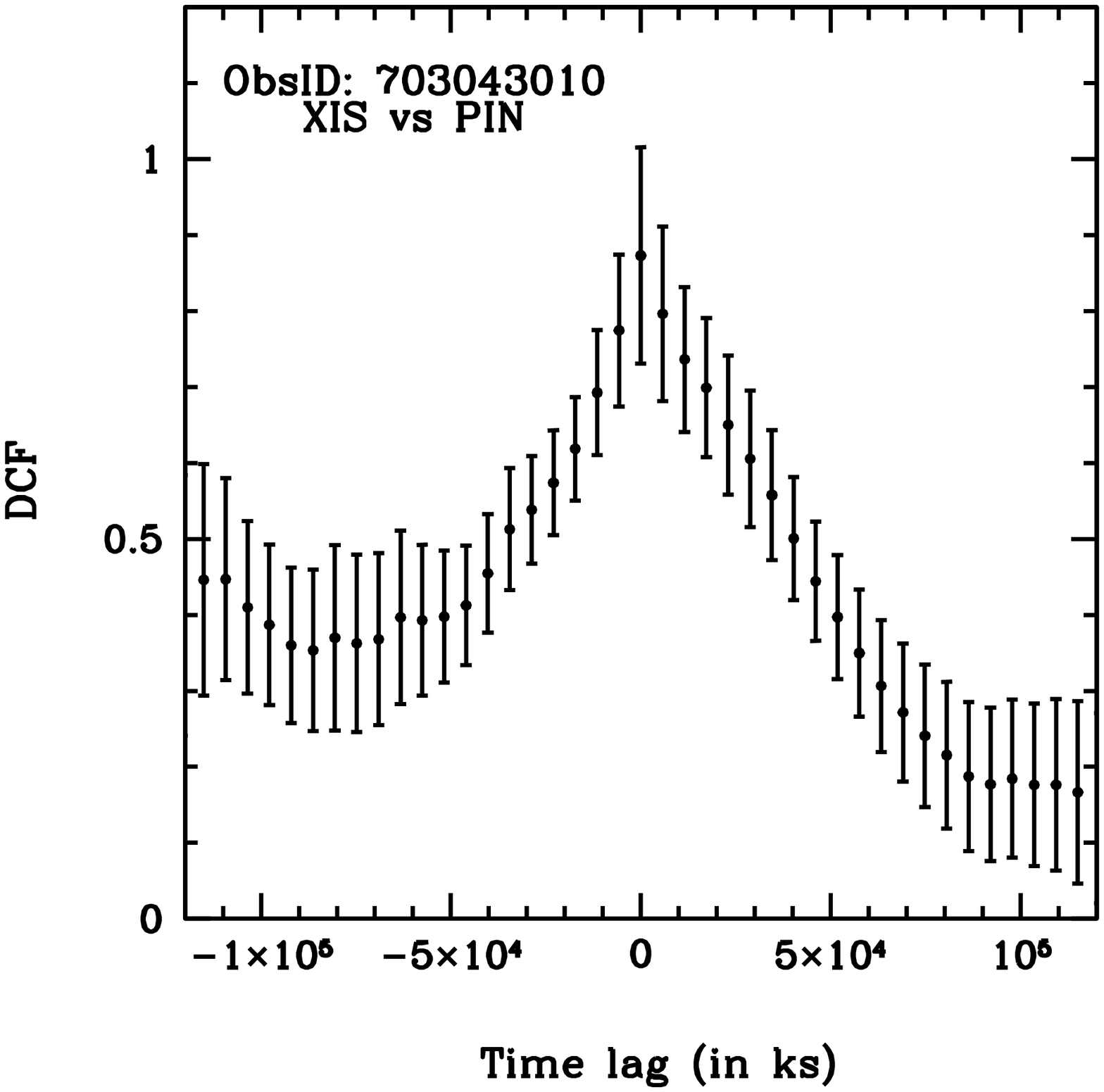}
\caption{ DCFs of X-ray LCs. Observation IDs and compared X-ray energy ranges  are given 
in each panel.}
\end{figure*}

\noindent
We computed Discrete Correlation Functions (DCFs) following \citet{Edelson1988}. This quantity is used to search for
possible variability time-scales and the time lags between multifrequency LCs.
The first step involves the calculation of the unbinned correlation (UDCF) using the given time series by:
\begin{equation}
UDCF_{ij} = {\frac{(a(i) - \bar{a})(b(j) - \bar{b})}{\sqrt{\sigma_a^2 \sigma_b^2}}}
\end{equation}
Here, $a(i)$ and $b(j)$ are the individual points in two time series $a$ and $b$, respectively, $\bar{a}$
and $\bar{b}$ are respectively the means of the time series, and $\sigma_a^2$ and $\sigma_b^2$ are their variances.
After the calculation of UDCF, this correlation function is binned to produce the DCF. Taking $\tau$ as the centre of a time bin and $n$
as the number of points in each bin, the DCF is found from the UDCF as:
\begin{equation}
DCF(\tau) = {\frac{1}{n}} \sum ~UDCF_{ij}(\tau) 
\end{equation}

\noindent
The  DCF analysis is used for finding the possible lags between the hard and soft X-ray bands. \\

\noindent
Since, the calculated DCF between various X-ray bands (shown in Figure 2) are broad, we fit them with a Gaussian function of the form:

\begin{equation}
DCF(\tau)=a \times {\rm exp}\Bigl[\frac{-(\tau - m)^{2}}{2 \sigma^{2}}\Bigr] 
\end{equation}

\noindent
Here, $a$ is the peak value of the DCF; $m$ is the time lag at which DCF peaks and $\sigma$ is the width of the Gaussian function.
The calculated parameters are presented in Table 3.

\noindent
\begin{table}
Table 3. Correlation analysis between X-ray bands
\begin{center}
\begin{tabular}{cccc} \hline \hline
ObsID              &   Bands                   &$m$ (ks) &$\sigma$ (ks)        \\\hline
701024010          & XIS soft vs. XIS hard     &~0.65$\pm$3.87   &10.34$\pm$3.87    \\
                   & XIS vs. PIN               &~0.89$\pm$3.95   &10.21$\pm$3.96  \\
703020010          & XIS soft vs. XIS hard     &~0.18$\pm$1.32   &16.48$\pm$1.33  \\
                   & XIS vs. PIN               &~8.32$\pm$3.40  &18.88$\pm$3.49  \\
703043010          & XIS soft vs. XIS hard     &~1.04$\pm$1.23   &33.36$\pm$1.65  \\
                   & XIS vs. PIN               &~2.74$\pm$1.17   &40.33$\pm$1.84 \\ \hline
\end{tabular}
\end{center}
\noindent
$m$= time lag at which DCF peaks \\
$\sigma$= width of the Gaussian function
\end{table}

\subsection{Power Spectral Density}

\noindent
A periodogram analysis produced by Fourier power spectral density (PSD) is a classical tool to search for 
the nature of temporal flux variations, including any possible periodicities and quasi-periodic oscillations 
(QPOs) in a LC. This method involves calculating the Fourier transform of the LC and 
then fitting the red noise variability of the PSD as a power-law. If the significance of any peak rising 
above the red noise is $3\sigma$ (99.73\%) or more, one normally considers it to provide a significant QPO 
detection. We followed the recipe given in \citet{Vaughan2005} to test for QPOs in the PSD. \\
\\
The PSD is calculated and normalization $N$ is defined such that the units of the periodogram will be
(rms/mean)$^{2}$/Hz. To fit the resultant red noise part of the spectrum $P(f)$ with respect to frequency $f$,
we assume a power law of form $P(f) = N~f^{\alpha}$,  where $N$ is the normalization constant and $\alpha$ is 
the power spectral index ($\alpha \leq$ 0) \citep{vanderKlis1989}. The significance levels are obtained by 
adding an appropriate term to the power spectrum.

\begin{figure*}
\begin{center}
\epsscale{0.95}
\includegraphics[width=5.9cm,angle=0]{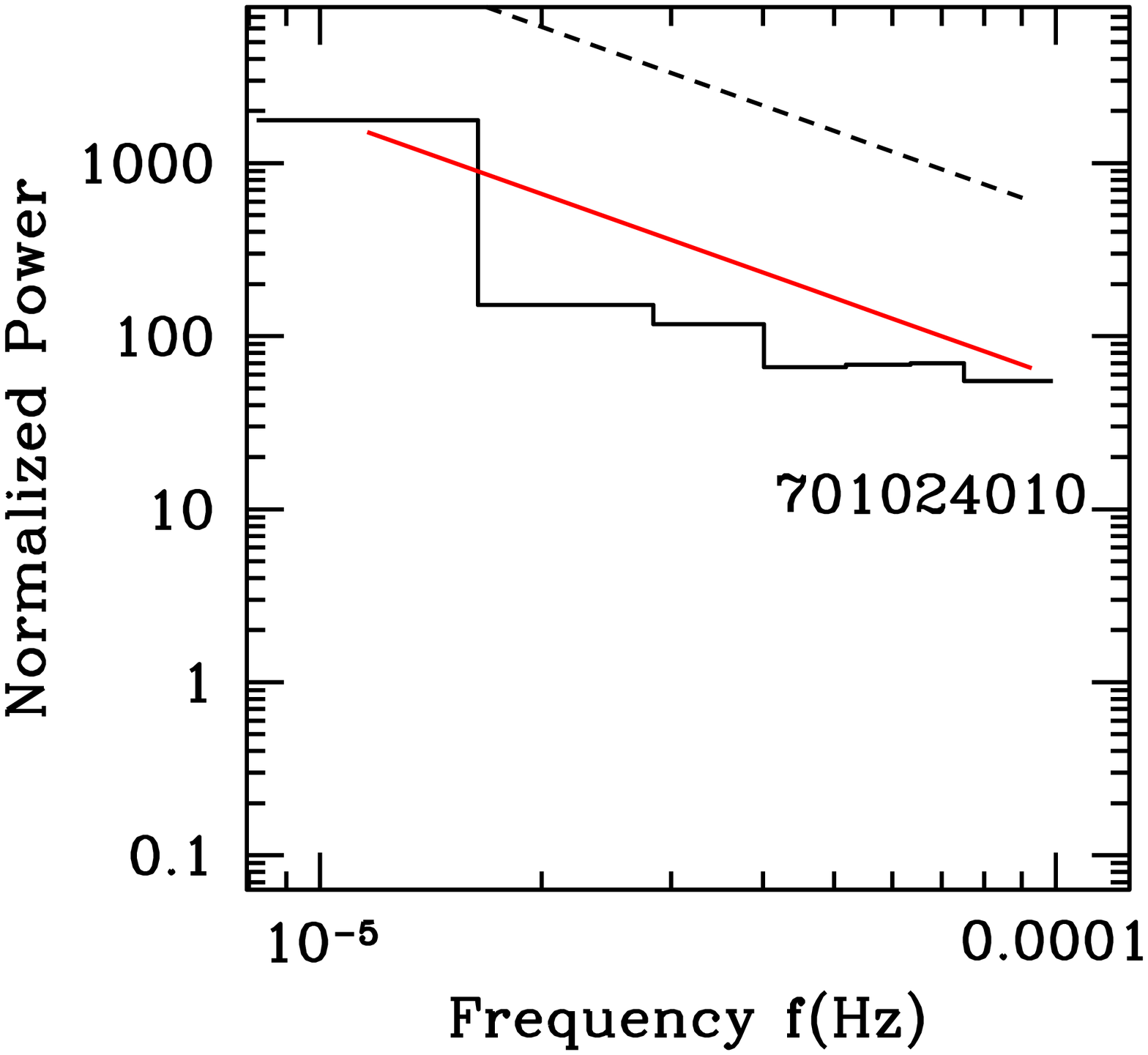} 
\includegraphics[width=5.9cm,angle=0]{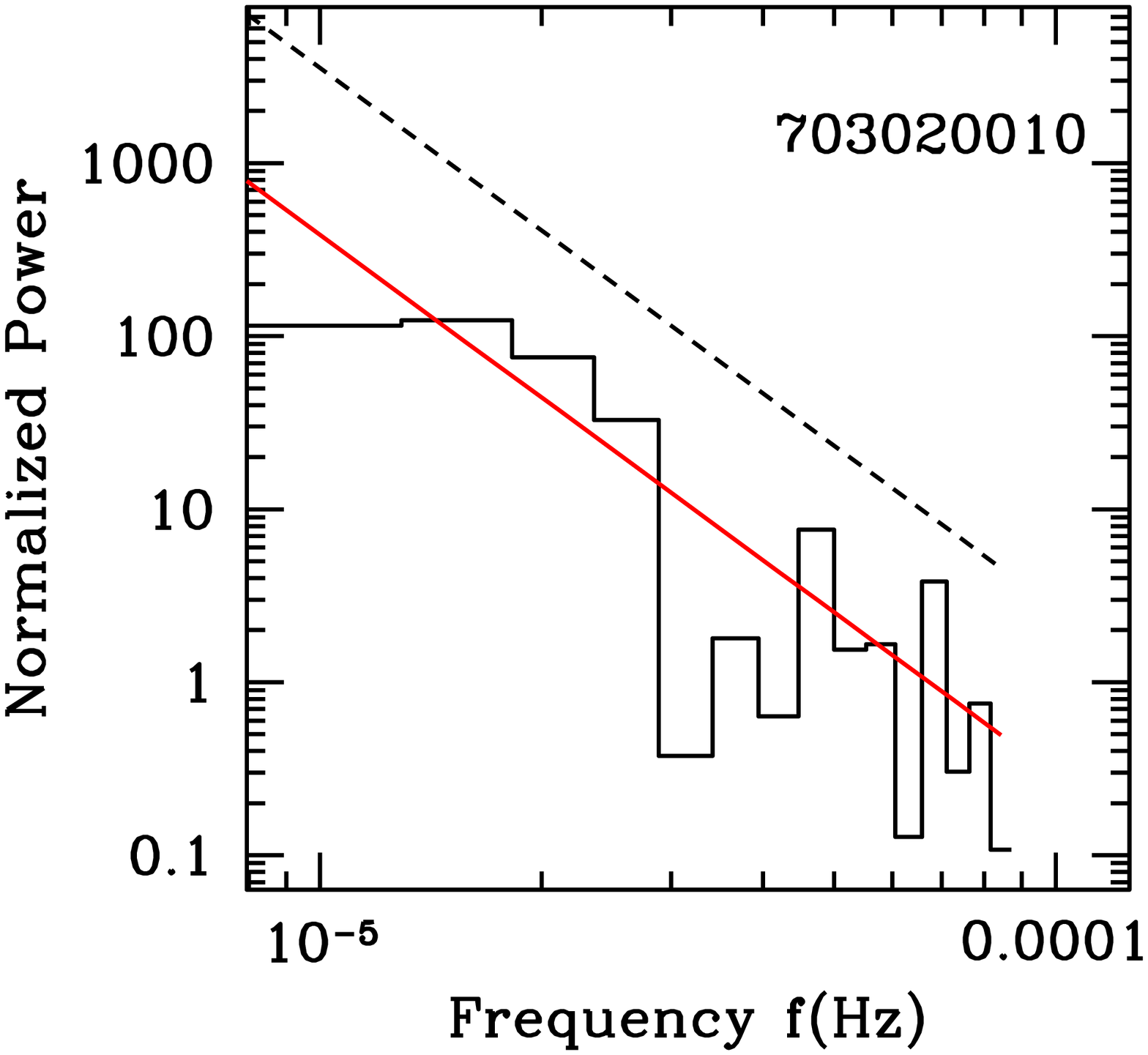}
\includegraphics[width=5.9cm,angle=0]{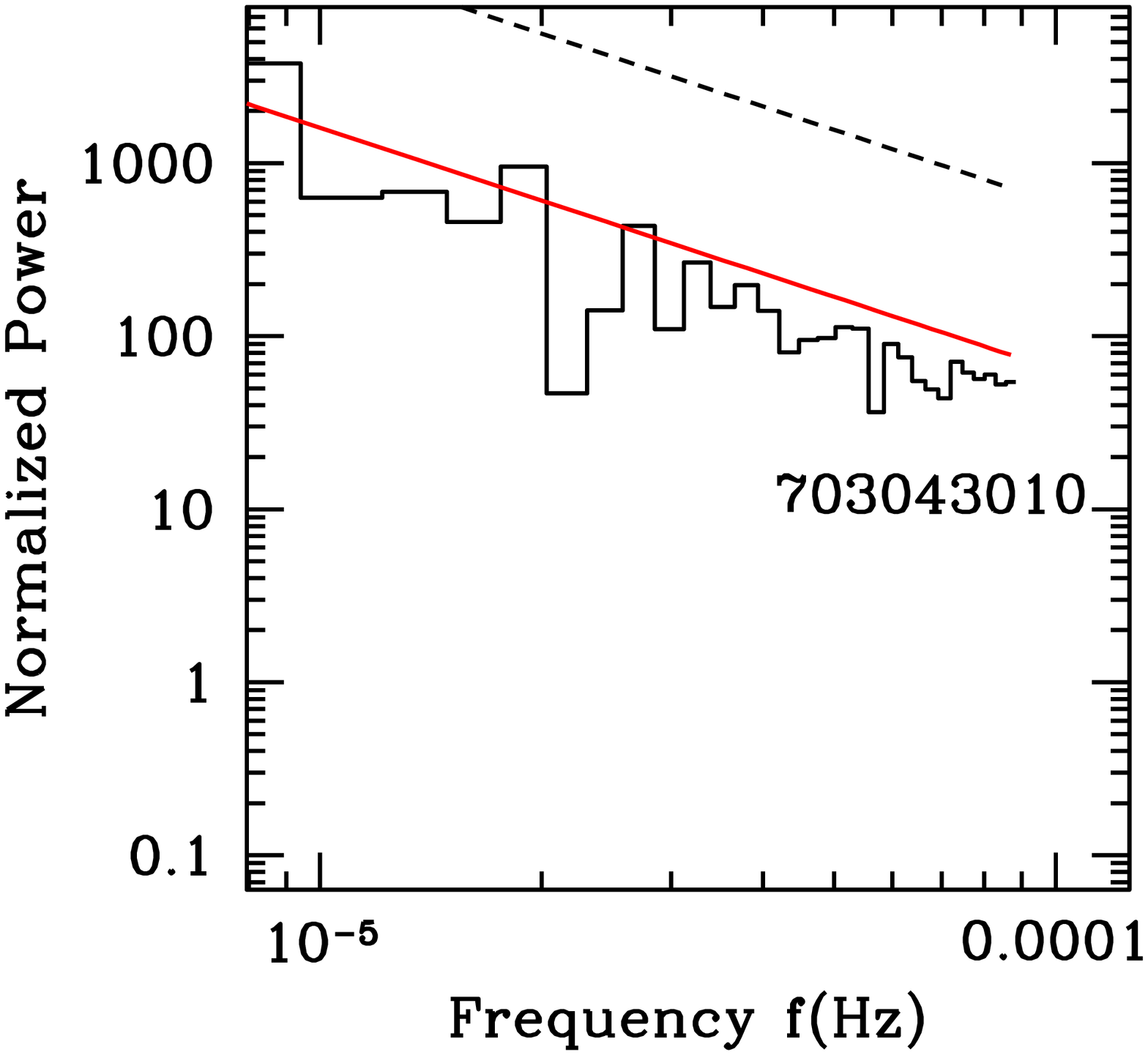} 
\figcaption{Power spectral densities (PSDs) of all XIS total (0.8 -- 8.0 keV) LCs of all three observations. 
Observation IDs are given in the PSD panels; the continuous red line is the red noise and the dotted black line shows the 99.73\% (3$\sigma$) confidence level for the red noise model.}  
\end{center}
\end{figure*}

\section{Results}

\noindent
We analyzed three publicly archived {\it Suzaku} observations of the blazar Mrk 421 which are spanned from $\sim$82 ks to $\sim$365 ks.
Details about the observations are in Table 1. The observation begun on 5 May 2008 (Obs ID: 703043010) which lasted 
for $\sim$ 365 ks is the longest, effectively  continuous and evenly sampled observation of any blazar in the X-ray energy 0.8 -- 60 keV 
which yielded a long LC to study the IDV of Mrk 421. The LCs we generated with these three observations used XIS (0.8 -- 1.5 keV (soft), 
1.5 -- 8 keV (hard), 0.8 -- 8 keV (total)) and PIN (12 -- 60 keV (total)) and are plotted in the top four panels of each figure 
of Figure 1. 

On visual inspection, the LCs of all three observation IDs in different energy bands show clear evidence of 
detection of IDVs. To quantify the IDV variability parameters, we estimated rms variability amplitude and its error using 
equations (5) and (6), respectively for all the LCs of these three observation IDs generated in different energy bands 
using the XIS and PIN detectors.  We also calculated the weighted variability timescales and their errors for XIS (total) and PIN (total) 
LCs using equations (7) and (9), respectively and the results are reported in Table 2. We noticed that the hard band 
1.5--8 of XIS and the total energy range 12--60 keV of PIN show the largest flux variations. The variations in the hard band 
1.5 -- 8 of XIS and for the total energy range 12 -- 60 keV of PIN for observation IDs 701024010, 703020010, and 703043010 are 
14.90$\pm$0.09\%, 11.75$\pm$0.07\%, 23.43$\pm$0.06\%; and 30.99$\pm$0.86\%, 20.62$\pm$0.02\%, 27.02$\pm$0.70\%, respectively.    \\    
\\
Hardness ratios are taken from the total band of the XIS detector as the soft band versus the total of the PIN detector as the hard band. We also produced a second HR by dividing the XIS output into soft and hard bands and both HRs are plotted with respect to time in lower two panels of Figure 1
for all three observations. Since we clearly noticed from  Table 2 that in the hard bands the 
variability is greater than in the soft bands, the HRs follow nearly the same trend as the LCs. The HR versus time 
plots in Figure 1 show that the source is harder when brighter and vice versa.  \\
\\
To determine the time lags between soft and hard X-ray bands, we computed DCFs between XIS (0.8 -- 1.5) keV (soft) versus XIS 
(1.5 -- 8) keV (hard), and XIS (0.8 -- 8) keV (soft) versus PIN (12 -- 60) keV (hard) data for all three observation IDs. All 
the DCF plots are presented in Figure 2. We fit all these DCF plots by using the Gaussian function equation (13) and the fitting 
parameters are provided in Table 3. It can be seen from  Table 3 that all of the time lags are consistent with being zero. 
Hence, we can say that we did not detect any significant lag between soft and hard bands observations. In the case of observation
ID 703020010, XIS (0.8 -- 8) keV (soft) versus PIN (12 -- 60) keV (hard) DCF plot show lag of 8.32$\pm$3.40 ks, but for this observation
the PIN data has two gaps in the data train and the uncertainty is quite large, so this can not be claimed as a genuine detection of a non-zero time lag. \\
\\
To attempt to characterize the temporal flux variations, and search for possible quasi-periodicity, we performed PSD analyses of the XIS total energy 
LCs of all three observations studied here. The PSD plots are presented in Figure 3. It is clear from this figure that the
PSDs of all three observations are red noise dominated and show no evidence of any quasi-periodicity. The slopes, $\alpha$, of this red noise are $-$1.51$\pm$0.27, 
$-$3.12$\pm$0.44, $-$1.40$\pm$0.11 with normalization constants $N$ of $-$4.28$\pm$1.19, $-$13.01$\pm$1.95, $-$3.79$\pm$0.51, 
respectively, for observation IDs 701024010, 703020010, 703043010. The average value of $\alpha$ = $-$2.01$\pm$0.53, but 
given the spread 
in values, this is not particularly relevant. Still, the slopes obtained here for the red noise are consistent 
with those computed for X-ray fluctuations seen in a wide range of AGN \citep{Gonzalez-Martin2012}.

\section{Discussion}

\noindent
In the present paper, we have employed three pointed essentially continuous observations of Mrk 421 which were taken by {\it Suzaku}
to study the X-ray IDV properties of this blazar in the energy range 0.8 -- 60 keV using the XIS and PIN detectors. Until now
these observations (IDs 701024010, 703020010, and 703043010) have not been used to study X-ray IDV properties of Mrk 421, so we 
are presenting the IDV results for the first time. The observation ID 701024010 begun on 28 April 2006 was studied by \citet{Ushio2009} 
where they looked for spectral evolution in 0.4 -- 60 keV and reported clear evidence of spectral variability.   
The second observation ID 703020010, begun on 13 December 2008, was analyzed with simultaneous {\it Swift} observations of the source
by \citet{Ushio2010}. They analyzed the spectrum with a parametric forward-fitting synchrotron model and with a particle
acceleration model and discussed the possibilities in the context of diffusive acceleration of electrons. The third observation,
ID 703043010, begun on 5 May 2008, was analyzed by \citet{Garson2010} and they found that a broken power-law model fit
the spectra very well. \\
\\
The study of flux variability on diverse timescales is an important tool to understand the emission mechanism both in blazars and 
in various other sub-classes of AGN. Rapid flux variability can be used to estimate the size, and constrain the location and structure of a dominant emitting 
region \citep[e.g.,][]{Ciprini2003}.  In blazars, most intrinsic flux variability across the EM bands can be explained by 
standard relativistic-jet-based models \citep[e.g.,][]{Marscher1985,GK1992,Marscher2014,Calafut2015} while for other AGN 
accretion disk based models should be most important 
\citep[e.g.,][]{Mangalam1993,Chakrabarti1993}. In the case of BL Lacs, the Doppler boosted jet emission dominates and if there is
any noticeable contribution by the accretion disk, it can be only seen when the blazar is observed in a low-flux state.
To detect an accretion disk contribution to changing blazar emission,  color variations in the time series data
and the presence of a big blue bump (BBB)  seen in the observed infrared to UV bands in the spectral energy distributions 
are necessary
\citep[e.g.,][and references therein]{Gu2006,Raiteri2007,Gaur2012b}. On the other hand, in the case of radio-quiet QSOs,
the flux variation on IDV and STV timescales can be explained by accretion disk based models, such as hot spots on or above the
disk or instabilities in the disk, perhaps driven by tilted disks or a dynamo \citep[e.g.,][and references therein]{Mangalam1993,Henisey2012,Sadowski2016}.  \\      
\\
Using these three observations by XIS and PIN detectors, we found that the shortest weighted variability timescale was 18.58 ks,
obtained from the Observation ID 701024010 with the PIN instrument in the 12--60 keV energy range. Now by adopting the 
simplest causality 
argument, the shortest variability timescale $\tau_{var}$ can be used to estimate the upper limit for the size of the 
emitting region, $R$,  as 

\begin{equation}
R\leq\frac{\delta }{\left ( 1+z \right )}c\tau_{var}
\end{equation}

\noindent
where, as usual, the Doppler factor $\delta =[\Gamma \left (1-\beta \cos \theta \right )]^{-1}$, with $\beta=v/c$ in terms of the bulk
velocity of the emitting region, $v$, and the bulk Lorentz factor $\Gamma=1/\sqrt{1-\beta^{2}}$. 
Here we have assumed that the emission originates from the a region of the blazar's jet which moves with relativistic 
speeds along the path that makes an angle, $\theta$, with the observer's line of sight. In the literature, the value of 
$\delta$ for Mrk 421 by using leptonic models ranges between 21 to 50 
\citep[e.g.,][and references therein]{Tavecchio1998,Abdo2011,Zhu2016}, whereas by using hadronic model, $\delta$ = 12 
was estimated \citep{Abdo2011}. 
By taking the shortest variability timescale 18.58 ks as 
mentioned above, 
and if we consider the complete range of Doppler factors ($\delta =$ 21 -- 50) for leptonic models, we estimate the upper limit 
of the size of the emitting region in the range of $\sim$ (1.1 -- 2.7) $\times$ 10$^{16}$ cm.  \\
\\

One of the fundamental quantities of an AGN one would like to know  is the mass of the central SMBH. 
The primary SMBH mass estimation methods include reverberation mapping and stellar or gas kinematics
\citep[e.g.,][]{Vestergaard2004}. Both of these methods are based on spectroscopy techniques and detected 
emission lines. Since Mrk 421 is a BL Lacertae object which has a featureless continuum, these primary methods 
are not directly applicable to determine the mass of its SMBH. Indirect estimation
of the mass of the central SMBH can be made through the spectra of the host galaxy of the BL Lacertae object if that
is discernable. 
Earlier this method was used to estimate the mass of the central SMBH of Mrk 421 which yielded a mass range 
(2 -- 9) $\times$ 10$^{8}$ M$_{\odot}$ \citep[e.g.,][]{Falomo2002,Wu2002,Barth2003,Treves2003}. \\
\\
In the case of active galaxies where the host galaxy is hard to observe beneath the dominant nuclear component, an alternative 
indirect method to estimate the SMBH mass involves fast 
variability timescales, where causality can limit size scales for active regions \citep{Gupta2012}. 
However, it is rare to  detect similar variability patterns, specifically doubling or exponentiating
timescales, in repeated observations of the same blazar at different times or in different EM bands. 
What variability timescales can be detected depend on the temporal resolution or cadence and total duration of the
observations as well as  the 
blazar flux state. If the fastest emission arises very close to the central SMBH 
(which is, however, more plausible for radio-quiet AGN than for blazars, where it likely to be in the jet and 
further from the SMBH), then this alternative method to estimate the upper limit of the mass of SMBH of Mrk 421 
can be used \citep{Gupta2012}. Explicitly, if we make the unlikely assumption that the X-rays detected by 
{\it Suzaku} were emitted close to the SMBH and not from the jet, and at around R = 5R$_{S}$, where R$_{S}$ = 2GM/c$^{2}$ 
is the Schwarzschild radius, the mass of SMBH can be approximated as
\begin{equation}
M_{BH} \approx {\frac {c^{3}t}{10G (1+z)}}.
\end{equation}

\noindent 
By using the shortest weighted variability timescale seen by {\it Suzaku} of 18.58 ks, one obtains a very rough mass estimate of the SMBH in
Mrk 421 of $\approx$ 4 $\times$ 10$^{8}$ M$_{\odot}$.   Recall that the  PIN weighted timescales from the other two observations are similar
but those from the XIS are longer.
But if the variability is due to the perturbations that were advected into the jet that arose from
the immediate region around the SMBH but are  
boosted along the observer's line of sight, then an additional Doppler boosting factor ($\delta$) is introduced in the
SMBH mass estimation \citep{Dai2007}.   Under these assumptions,  for $\delta =$ 21 and 50, the 
estimated mass of the SMBH in Mrk 421 would be $\sim 8 \times 10^{9}$ and $\sim 2 \times 10^{10}$ M$_{\odot}$, respectively.
As noted above, earlier attempts at estimating the SMBH mass of  Mrk 421 yielded a mass range 
(2 -- 9) $\times$ 10$^{8}$ M$_{\odot}$ \citep[e.g.,][]{Falomo2002,Wu2002,Barth2003,Treves2003}. 
So the very crude mass estimation of Mrk 421 for variations arising from the immediate vicinity of the SMBH case is consistent 
with those, while those incorporating Doppler boosting of such perturbations are not. \\
\\
For estimation of other parameters in models where the emission is predominantly from the
jets, and has no direct connection with the near vicinity of the SMBH, we now adopt a moderate value of $\delta = 25$, as also recently used for Mrk 421 by
others \citep[e.g.,][]{Balokovic2016,Pandey2017,Aggrawal2018}.  
In the comoving frame, a diffusive shock acceleration mechanism is often assumed to be responsible for electron acceleration 
in blazar jets \citep[e.g.,][]{Drury1983,Blandford1987}. For the diffusive shock acceleration mechanism, \citet{Zhang2002} 
has given the acceleration timescale of electron with energy ${\it E = \gamma m_{e}c^{2}}$ as  

\begin{equation}
t_{acc}(\gamma) \simeq 3.79\times10^{-7} \frac{(1+z)}{\delta} \xi B^{-1}\gamma ~{\rm s},
\end{equation}

\noindent
where $\xi$ is the acceleration parameter, $B$ is the magnetic field in Gauss, and $\gamma$ is the Lorentz factor of 
the ultrarelativistic electrons. \\
\\
Mrk 421 is a TeV blazar and belongs to the high energy blazar (HBL) sub-class, and is also known as a member of 
the high synchrotron peak (HSP)
blazar class\footnote{http://tevcat.uchicago.edu} in which X-ray emission is mainly synchrotron radiation. The 
synchrotron cooling timescale of a relativistic electron with ${\it E = \gamma m_{e}c^{2}}$, t$_{cool}(\gamma)$
\citep[see, e.g.,][]{Rybicki1979} is given as  

\begin{equation}
t_{cool}(\gamma) \simeq 7.74\times10^{8}\frac{(1+z)}{\delta} B^{-2}\gamma^{-1} ~{\rm s}.
\end{equation} 

\noindent
For the {\it Suzaku} total energy range of 0.8--60 KeV, the critical synchrotron emission frequency is

\begin{equation} 
\nu \simeq 4.2 \times 10^{6} {\frac{\delta}{1+z}} B \gamma^{2} \simeq 10^{19} \nu_{19} ~{\rm Hz}
\end{equation}

\noindent
The shortest weighted variability timescale we saw for Mrk 421 is 18.58 ks, where $F_{var}$ $\sim$ 31\%. The cooling timescale 
should be longer than or equivalent to this minimum variability timescale, which implies

\begin{equation} 
B \geq 0.21(1+z)^{1/3}\delta^{-1/3}\nu^{-1/3}_{19} ~{\rm G}.
\end{equation}

\noindent
For the value of $\delta$=25 we have adopted one gets
\begin{equation}
B \geq 0.07~ \nu^{-1/3}_{19} {\rm G}.
\end{equation}

\noindent
An estimate of  $B$ $\leq$ 0.1 G was the typical value for Mrk 421 obtained using SED modeling \citep{Paliya2015}.
Similar values of $B$ were also obtained for Mrk 421 using {\it NuStar} and {\it Chandra} observations 
\citep{Pandey2017,Aggrawal2018}. Combining these consistent values of $B$ and $\delta$, we estimate the electron Lorentz factor as

\begin{equation}
\gamma \geq 1.4\times10^{6} ~\nu^{2/3}_{19}.
\end{equation}

\noindent
Hard and variable X-ray emission can be directly attributed to the relativistic electrons in TeV HSP blazars.  
We investigated this emission by using {\it Suzaku} data in a broad X-ray band from 0.8 -- 60 keV, and the short cooling timescales of these  
high-energy relativistic electrons \citep{Pandey2017} imply that  the acceleration process takes place repeatedly. 
Diffusive-shock acceleration \citep{Blandford1987} could  be responsible for both spectral hardening and 
variations in the flux at high energies. \\ 
\\
For the HSP blazar Mrk 421 we studied X-ray spectral variability by analyzing two hardness ratios (HRs),
namely, 0.8--1.5 keV versus 1.5--8 keV, and 0.8--8 keV versus 12--60 keV.  Although the HR is an  easy and 
efficient way to study the change in X-ray spectra, it is crude, and we cannot directly estimate the physical parameters 
which are responsible for spectral variability by this method. We noticed that the both HR plots for all three 
observation IDs show patterns similar to those of the LCs, which shows that the source has relatively larger 
amplitude variations in harder X-rays than in softer ones (see Fig.\ 1). Hence, it seems 
that the {\it Suzaku} observations of Mrk 421 presented in the present work follow the general trend of 
``harder-when-brighter" which generally characterizes the HSP type blazars 
\citep[e.g.,][and references therein]{Pandey2017,Aggrawal2018}. 

\section{Conclusions}

\noindent
We studied the three {\it Suzaku} light curves observations of the TeV HSP blazar Mrk 421 which are available 
in its public archive. These observations were all those taken of this source during complete operational span of the 
satellite. We searched for IDV and its time scales, hardness ratios, energy lags between soft and hard energies, and also performed PSD
analyses to characterize the IDV and search for any possible QPO present. Our conclusions are summarized 
as follows: \\
\\
$\bullet$ The fractional variability clearly shows that the source shows large amplitude IDV for all three
observation IDs in all soft and hard bands of both the instruments (XIS and PIN) on board {\it Suzaku}. \\
\\
$\bullet$ The fractional variability is lower in the soft bands than in the hard bands. We estimated the IDV
timescale for all three observation IDs in 0.8 -- 8 keV (XIS total) and 12 -- 60 keV (PIN total) and the weighted
variability timescales are found in the range of 18.58 ks to 78.07 ks. The shortest IDV timescale 18.58 ks was 
used to estimate the various parameters of the blazar emission. \\
\\
$\bullet$ Using the DCF method, we estimated lags between 0.8--1.5 keV (soft) versus 1.5--8 keV (hard),
and 0.8--8 keV (soft) versus 12--60 keV (hard) for all three observations. All the DCF peaks are consistent with
zero lag within a general broader Gaussian profile. This supports the hypothesis that the emission in 
these different X-ray bands are co-spatial
and are produced by the same population of leptons. \\
\\
$\bullet$ Our HR analysis  for 0.8--1.5 keV (soft) versus 1.5--8 keV (hard) and 0.8--8 keV (soft) 
versus 12--60 keV (hard)  show similar patterns as the light curves of all three observations.
This implies that hard bands are more variable than the soft bands. This source exhibits the general harder-when-brighter
behavior of HSP blazars . \\
\\
$\bullet$ A PSD analysis for each of the three observations of the XIS total energy data was performed.  These PSDs  
 are red noise dominated, with slopes ranging from $-1.4$ to $-3.1$,  and there is no significant peak that might indicate a possible 
QPO.  \\
\\
$\bullet$ Under the unlikely assumption that the fastest variations detected correspond to the region close to the SMBH,
its mass in Mrk 421 is  estimated to be $\sim$ 4 $\times$ 10$^{8}$ M$_{\odot}$. By assuming the variability is due to perturbations 
arising close to the SMBH but moving into the jet and thus boosted along the observer's line of sight, the mass in Mrk 421 could be a factor of 20--50 higher.

\section*{ACKNOWLEDGMENTS}
\noindent
This research has made use of data obtained from the {\it Suzaku} satellite, a collaborative mission between 
the space agencies of Japan (JAXA) and the USA (NASA). \\  
\\
We thank the anonymous referee for useful comments which helped us to improve the manuscript. ACG thanks 
Dr. Gopal Bhatta of Astronomical Observatory of the Jagiellonian University, Krakow, Poland for discussions 
about estimation of the shortest variability timescale and for providing his computer code. \\ 
\\
This work is funded by the National Key R\&D Programme of China (under grant Nos. 2018YFA0404602 and 2018YFA0404603) 
and Chinese Academy of Sciences (under grant No. 114231KYSB20170003). ZZL is thankful for support from the Chinese 
Academy of Sciences (CAS) Pioneer Hundred Talent Program. ACG is grateful for hospitality at SHAO, Shanghai, China and 
Astronomical Observatory of the Jagiellonian University, Krakow, Poland, where this paper was written. The work of ACG 
is partially supported by Indo-Poland project No.\ DST/INT/POL/P19/2016 funded by Department of Science and Technology (DST), 
Government of India, and also partially supported by CAS President's International Fellowship Initiative 
(PIFI) grant no.\ 2016VMB073. HG acknowledges financial support from the Department of Science \& Technology 
(DST), Government of India through INSPIRE faculty award IFA17-PH197 at ARIES, Nainital. MFG is supported by the 
National Science Foundation of China (grants 11873073, U1531245 and 11473054). HGX is supported by the Ministry of 
Science and Technology of China (grant No. 2018YFA0404601), and the National Science Foundation of China 
(grant Nos. 11433002, 11621303 and 11973033).

\software{HEAsoft \citep[v6.17; HEASARC 2014;][]{Blackburn1995}}

\clearpage

\end{document}